\pdfoutput=1
\documentclass{article} 

\usepackage{colm2024_conference}

\colmfinalcopy

\usepackage[scaled=.90]{helvet}
\usepackage{courier}
\usepackage[hidelinks]{hyperref}

\usepackage{latexsym}
\usepackage[T1]{fontenc}
\usepackage[utf8]{inputenc}
\usepackage{microtype}

\usepackage{amsmath}
\usepackage{amssymb}
\usepackage{bbold}
\usepackage{multicol}

\usepackage[inline]{enumitem}
\usepackage{enumerate}
\usepackage{mathtools}

\usepackage{booktabs,caption}
\usepackage[flushleft]{threeparttable}
\usepackage[most]{tcolorbox}

\usepackage{soul}
\usepackage{dsfont}
\usepackage{color, colortbl}
\usepackage[most]{tcolorbox}
\definecolor{LightCyan}{rgb}{0.88,1,1}
\definecolor{Gray}{gray}{0.9}

\newcommand{\code}[1]{{\ttfamily#1}}
\newcommand{\header}[1]{\vspace*{1mm}\noindent\textbf{#1}}

\usepackage{bbold}

\begin{document}
\title{Instruction Distillation Makes Large Language Models\\
Efficient Zero-shot Rankers}

\author{Weiwei Sun\textsuperscript{\rm 1} \quad Zheng Chen\textsuperscript{\rm 1} \quad Xinyu Ma\textsuperscript{\rm 2} \quad Lingyong Yan\textsuperscript{\rm 2} \quad \textbf{Shuaiqiang Wang}\textsuperscript{\rm 2} \\
\textbf{Pengjie Ren}\textsuperscript{\rm 1} \quad \textbf{Zhumin Chen}\textsuperscript{\rm 1} \quad \textbf{Dawei Yin}\textsuperscript{\rm 2} \quad \textbf{Zhaochun Ren}\textsuperscript{\rm 3} \\
\textsuperscript{\rm 1}Shandong University, Qingdao, China \quad
\textsuperscript{\rm 2}Baidu Inc., Beijing, China \\
\textsuperscript{\rm 3}Leiden University, Leiden, The Netherlands\\
\texttt{\{sunnweiwei,xinyuma2016,lingyongy\}@gmail.com}\\
~\texttt{yindawei@acm.org,}~\texttt{z.ren@liacs.leidenuniv.nl}
}

\maketitle

\begin{abstract}
Recent studies have demonstrated the great potential of Large Language Models (LLMs) serving as zero-shot relevance rankers.
The typical approach involves making comparisons between pairs or lists of documents.
Although effective, these \textit{listwise} and \textit{pairwise} methods are not efficient and also heavily rely on intricate prompt engineering.
To tackle this problem, we introduce a novel instruction distillation method.
The key idea is to distill the pairwise ranking ability of open-sourced LLMs to a simpler but more efficient \textit{pointwise} ranking. 
Specifically, given the same LLM, we first rank documents using the effective pairwise approach with complex instructions, and then distill the teacher predictions to the pointwise approach with simpler instructions.
Evaluation results on the BEIR, TREC, and ReDial datasets demonstrate that instruction distillation can improve efficiency by 10 to 100$\times$ and also enhance the ranking performance of LLMs. 
Furthermore, our approach surpasses the performance of existing supervised methods like monoT5 and is on par with the state-of-the-art zero-shot methods.
The code to reproduce our results is available at \url{www.github.com/sunnweiwei/RankGPT}.

\end{abstract}

\section{Introduction}

Large Language Models (LLMs), such as ChatGPT and GPT-4, have achieved remarkable success in various Natural Language Processing (NLP) tasks~\citep{ChatGPT, OpenAI2023GPT4TR}. 
One notable capability of LLMs is their ability to solve tasks using carefully designed prompts or instructions~\citep{NewBing}. 
This has drawn much attention from the Information Retrieval (IR) community given its potential to significantly reduce the huge annotation costs~\citep{Shi2023REPLUGRB, Sun2023IsCG}.

Relevance ranking has been the most critical problem in IR, which aims at ranking a set of candidate items by their relevance given the query~\citep{Fan2021PretrainingMI}.
Recently, there has been a series of works using large models as zero-shot rankers through pointwise, pairwise, and listwise ranking prompting, and these have achieved impressive results on IR benchmarks~\citep{Sun2023IsCG,Ma2023ZeroShotLD,Qin2023LargeLM}.

\begin{figure}[!t]
\centering
\includegraphics[width=0.6\columnwidth]{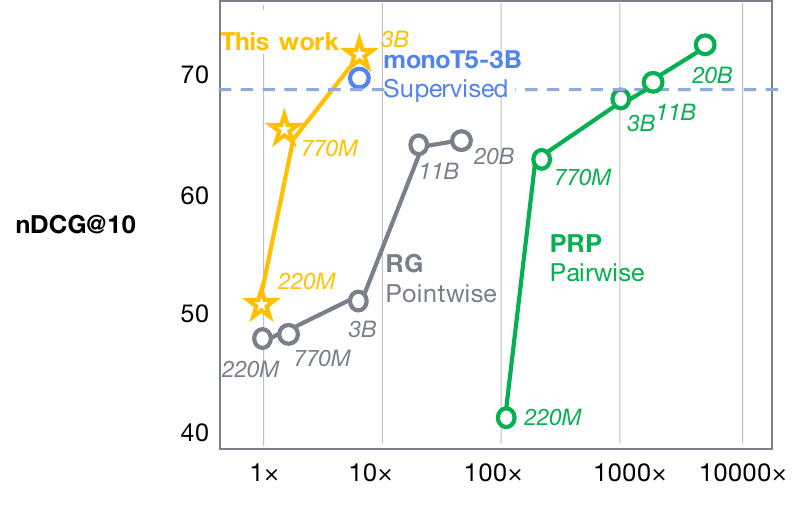} 
\caption{The average nDCG@10 of various LLM-based re-ranking methods on TREC benchmarks. The horizontal axis represents the speed of each method relative to monoT5-Base~\citep{Nogueira2020DocumentRW}, as measured by the average latency time per query. All methods are based on the T5 series foundation models. RG refers to the relevance generation method, and PRP refers to the pairwise ranking method.
}
\label{fig:speed}
\end{figure}

Employing LLMs for ranking tasks still faces several practical challenges, including application efficiency and output stability.
On one hand, both listwise and pairwise ranking methods suffer from efficiency issues. For listwise ranking~\citep{Sun2023IsCG,Ma2023ZeroShotLD}, the exponential time complexity of the Transformer with respect to input length renders it impractical for many industrial applications. Pairwise ranking requires pairing every document with every other, with the obvious drawback being its costly $O(n^2)$ calls to LLMs~\citep{Qin2023LargeLM}. On the other hand, while pointwise ranking is more efficient, it compromises on effectiveness~\citep{Liang2022HolisticEO}. The pretraining objective of LLMs isn't inherently tailored for ranking tasks (i.e., generative language modeling vs. relevance ranking), meaning its prediction probability isn't calibrated to the relevance score~\citep{Zhao2021CalibrateBU,zhao2023knowing}. Other challenges, such as unstable outputs, position bias, and repetitions from LLMs, become more pronounced in IR tasks, where deterministic output in terms of relevance is crucial~\citep{Sun2023IsCG}.



To address these challenges, this paper introduces a novel \emph{Instruction Distillation} method to enhance the efficiency and stability of LLMs in the ranking task.
The key idea is to distill the predictions of pairwise ranking (PRP) with computationally demanding instruction (\textit{teacher instruction}) to the efficient pointwise prompting method but with simpler instruction (\textit{student instruction}). 
Through this distillation process, the task instructions used for ranking are substantially simplified, leading not only to increased efficiency but also to enhanced performance.
In this work, we use open-sourced LLMs FLAN-T5 and our method is zero-shot text ranking since FLAN-T5 is not directly exposed to human-labeled data.


We empirically evaluate instruction distilled models against other baselines in Figure~\ref{fig:speed}.
These distilled student models are between 10 and 100$\times$ more efficient compared to their teacher models (i.e., PRP) while also yielding significant enhancements.
Compared to vanilla pointwise ranking methods (Relevance Generation methods, RG), our distilled models show a $40\%$ performance improvement in terms of nDCG@10. 
Remarkably, our distilled FLAN-T5-XL model even surpasses the SOTA supervised systems like monoT5-3B~\citep{Nogueira2020DocumentRW} in IR benchmarks. 
This is particularly notable as it achieves this without relying on any human relevance judgments. 
We also condu
Further verification is conducted on various ranking tasks such as the BEIR benchmark and the conversational recommendation tasks present in the REDIAL benchmark.

In summary, this paper makes the following contributions:
\begin{itemize}
    \item We propose Instruction Distillation, an unsupervised approach to specialize LLMs on IR tasks by distilling instructions.
    \item We show the instruction distilled LLM is both more efficient and effective compared to existing zero-shot LLMs with the same amount of parameters. 
    \item We illustrate the robust performance of our method in both passage ranking and movie recommendation tasks, surpassing the state-of-the-art supervised methods.\footnote{Code and pre-trained models are available at \url{https://github.com/sunnweiwei/RankGPT/tree/main/InstructDistill}}
\end{itemize}

\section{Related Work}

\subsection{LLMs for Information Retrieval}
Large language models (LLMs) have been pre-trained on a large-scale corpus and possess strong text understanding and reasoning capabilities~\citep{OpenAI2023GPT4TR, Anil2023PaLM2T, Shoeybi2019MegatronLMTM, Touvron2023LLaMAOA}.
Recently, LLMs have found increasing applications in information retrieval~\citep{Zhu2023LargeLM, Wu2023ASO, Yu2022GenerateRT,Sun2023GenerativeKS,Hou2023LargeLM,Sun2023LearningTT,Bao2023TALLRecAE}.
These methods can be broadly divided into two categories: synthetic data generation and relevance ranking.

Several approaches have been proposed to utilize LLMs to generate synthetic data for IR. 
For example, SGPT~\citep{Muennighoff2022SGPTGS} generates text embeddings using GPT for dense retrieval; and \citet{Gao2022PreciseZD,Wang2023Query2docQE} proposes to generate pseudo-documents using LLMs and retrieve these pseudo-documents first using queries.
\citet{Dai2022PromptagatorFD} proposes to generate pseudo-queries for few-shot dense retrieval.

In addition, LLMs have also been used for relevance ranking tasks. UPR~\citep{Sachan2022ImprovingPR} and SGPT-CE~\citep{Muennighoff2022SGPTGS} introduce instructional query generation methods, which rank documents based on the generation likelihood of query given the document.
HELM~\citep{Liang2022HolisticEO} utilizes instructional relevance generation for ranking, prompting LLMs to generate relevance proxy tokens and rank documents based on the generation probability. 
RankGPT~\citep{Sun2023IsCG} proposes a zero-shot permutation generation method, which prompts LLMs to directly generation the ranking permutation and its performance surpasses supervised models when based on GPT4.
\citet{Qin2023LargeLM} proposes a pairwise ranking prompting method (PRP) based on open-sourced LLMs.

%
Though good results are achieved by the methods above, two challenges still remain:
(1) Unstable output, sensitivity of input, repetition, and position bias could harm the performance severely.
(2) Sophisticated instruction techniques and task designs are commonly adapted to achieve high performance at the cost of computational complexity. It would be hard for computationally costly methods to be applied to a practical scenario.

\subsection{LLMs Distillation}
Despite their impressive capabilities, LLMs such as GPT-4 often come with high costs and lack open-source availability. As a result, considerable research has explored various ways to distill the capabilities of LLMs into specialized, customized models. For instance, \citet{Fu2023SpecializingSL} and \citet{Magister2022TeachingSL} have successfully distilled the reasoning ability of LLMs into smaller models. Self-instruct~\citep{Wang2022SelfInstructAL, alpaca} propose iterative approaches to distill GPT-3 using their outputs. 

Additionally, \citet{Sachan2022QuestionsAA} and \citet{Shi2023REPLUGRB} utilize the generation probability of LLMs to improve retrieval systems.
\citet{Snell2022LearningBD} introduces a similar context distillation method to simplify the overlong context when prompting LLMs on Text-to-SQL tasks.
This paper presents the Instruction Distillation method, aiming at distilling the ability explored by sophisticated instructions into the model using more efficient instructions to enhance the model efficiency and output stability. 

\section{Method}

In this section, we introduce the instruction distillation method in detail. This novel approach enhances both the effectiveness and efficiency of open-sourced LLMs during the inference stage by distilling the capabilities harnessed by complex instructions into a more efficient one. Thus, when deploying to real-world applications, our methodology is able to obtain good performance which necessitates only lower computation costs compared to others.

\subsection{Task Formalization}
The task of \textit{relevance ranking} can be formally defined as follows: Given a query \(q\) and a set of candidate items \(D = \{d_1, \ldots, d_n\}\), the objective is to determine the ranking of these candidates, represented as \(R = \{r_1, \ldots, r_n\}\). Here, \(r_i \in \{1, 2, \ldots, n\}\) denotes the rank of candidate \(d_i\). For instance, if \(r_i = 3\), it denotes that \(d_i\) is ranked third among the \(n\) candidates.

A ranking model, denoted as \(f(\cdot)\), assigns scores to the candidates based on their relevance to the query:
\begin{equation}
    s_i = f(q, d_i)
\end{equation}
Subsequently, the candidates are ranked according to these relevance scores: $r_i = \text{arg sort}_i (s_1, \ldots, s_n)$

\subsection{Prompting LLMs for Ranking Tasks}
Recent studies have explored the potential of using Large Language Models (LLMs) for the re-ranking task. 
Diverse prompting strategies have been explored.
Based on the type of instruction employed, existing strategies can be categorized into three types: (1) pointwise ranking, (2) pairwise ranking, and (3) listwise ranking~\citep{Wu2023ASO, Zhu2023LargeLM}.

\header{Pointwise Ranking} assigns an independent score to each item $d_i$, subsequently ranking the set $D$ based on these scores.
A prevalent pointwise prompting approach for LLMs is instructional relevance generation, which is exemplified in HELM~\citep{Liang2022HolisticEO}.
In this approach, LLMs are prompted to output either "Yes" or "No" to determine the relevance of the candidates to a given query. The generation probability is then converted to the relevance score:
\begin{equation}
s_i = \begin{cases} 
1 + f(\text{Yes}\mid \mathcal{I}_{\text{RG}}(q, d_i)), &\text{if output Yes}\\
1 - f(\text{No}\mid \mathcal{I}_{\text{RG}}(q, d_i)), &\text{if output No}
\end{cases}
\end{equation}
Here $f(\cdot)$  represents the large language model, and $\mathcal{I}_{\text{RG}}$ denotes the relevance generation instruction that converts the input $q$ and $d_i$ into the test-based prompt.

\begin{equation}
    s_i = \frac{1}{|q|} \sum_{t} \log p(q_t\mid q_{<t}, p_i, \mathcal{I}_{\text{query}})
\end{equation}

\header{Pairwise Ranking} is employed by PRP~\citep{Qin2023LargeLM}. In this technique, both the query and a pair of candidate items serve as prompts, guiding the LLMs in ranking tasks. 
For every pair of items $d_i$ and $d_j$, a specific pairwise comparison instruction, denoted by $\mathcal{I}_{\text{PRP}}$, is employed to instruct the LLMs, i.e., $f(\cdot)$, to determine which item is more relevant to the given query. This can be formalized as:
\begin{equation}
c_{i,j} = \begin{cases} 
1, &\text{if}~f(\mathcal{I}_{\text{PRP}}(q, d_i, d_j)) = i\\
0, &\text{if}~f(\mathcal{I}_{\text{PRP}}(q, d_i, d_j)) = j\\
0.5, &\text{else}
\end{cases}
\end{equation}
Here, $c_{i,j}$ denotes the LLM's choice. Considering that LLMs may exhibit sensitivity to the order of text in the prompt, for every pair $d_i$ and $d_j$, PRP consults the LLM twice, inverting their order between $\mathcal{I}_{\text{PRP}}(q, d_i, d_j)$ and $\mathcal{I}_{\text{PRP}}(q, d_j, d_i)$.
Subsequently, to compute the relevance score of the $i$-th candidate $d_i$, PRP compares $d_i$ against all other candidates in the set $D$:
\begin{equation}\label{eq:pair}
    s_i = \sum_{j \neq i} c_{i,j} + (1 - c_{j,i})
\end{equation}
The final relevance score aggregates all comparison results.


\header{Listwise Ranking} has been adopted by \citet{Sun2023IsCG, Ma2023ZeroShotLD}. This approach involves feeding a set of items into the LLMs, where each item is identified by a unique identifier (e.g., [1], [2], etc.). The LLMs are then instructed to generate a permutation of these items, such as ``[2] > [3] > [1] > \dots'':
\begin{equation}
    \mathbf{Perm} = f(\mathcal{I}_{\text{List}}(q, d_1, d_2, \ldots, d_n))
\end{equation}
This generated permutation \( \mathbf{Perm} \) can be readily transformed into ranking results \( R \), which bypasses the necessity to compute an explicit relevance score, \( s_i \), for each candidate \( d_i \). 
To ensure consistency in notation with scoring-based methodologies, the relevance score \( s_i \) is defined as the reciprocal of its rank: \( s_i \coloneqq \frac{1}{r_i} \).

\begin{table}[!t]
\centering
\small
\setlength\tabcolsep{3pt}
\caption{Computational complexity of different instruction methods. $n$ is the number of items to be ranked. $k$ is a constant related to the sliding window method.}
\label{table:complexity}
\begin{tabular}{lcc}
\toprule
Instruction & Complexity & Examples\\
\midrule
Pointwise Ranking & $O(n)$ & \citep{Liang2022HolisticEO,Sachan2022ImprovingPR} \\
Pairwise Ranking & $O(n^2)$ & \citep{Qin2023LargeLM} \\
Listwise Ranking & $O(k*n)$ & \citep{Sun2023IsCG,Ma2023ZeroShotLD} \
\\
\bottomrule

\end{tabular}
\end{table}

\subsection{Computational Complexity of Different Instructions.}

Different ranking instructions offer various trade-offs in terms of efficiency and effectiveness. 
A summary of these instructions is listed in Table~\ref{table:complexity}.
Among these, the pointwise ranking is computationally the most efficient, having a complexity of \(O(N)\). Nevertheless, this approach requires the model to yield a calibrated pointwise score, a feat which is notably challenging.

In contrast, the pairwise ranking paradigm resolves the calibration issue by engaging in one-to-one pairwise comparisons. This solution, however, elevates the computational complexity to \(O(N^2)\). To tackle this, \citet{Qin2023LargeLM} propose two methods to curtail the pairwise ranking's complexity: sorting and the sliding window technique. While promising, these methods are still in their nascent stages, proving challenging to stabilize and parallelize.

On another note, listwise ranking demonstrates good performance when tested on commercial and also proprietary LLMs, such as GPT-4. However, it performs poorly on smaller, open-source models. A possible reason could be the inferior comprehension of instructions in these open-source counterparts.

In summary, each ranking method comes with its set of pros and cons: the pointwise approach is efficient but may not be highly effective; the pairwise method is effective but computationally demanding; and the listwise method is most effective but limited to closed-source LLMs like GPT-4. These insights set the stage for our novel solution -- the \textit{instruction distillation} strategy., which we will introduce in the next section. 

An overview of the proposed instruction distillation approach is presented. Instruction distillation distills the abilities obtained from complex instruction techniques (e.g., pairwise ranking) into a model that is more efficient with simple instruction techniques (e.g., pointwise ranking).

\subsection{Instruction Distillation}
\begin{figure}[!t]
\centering
\includegraphics[width=0.85\columnwidth]{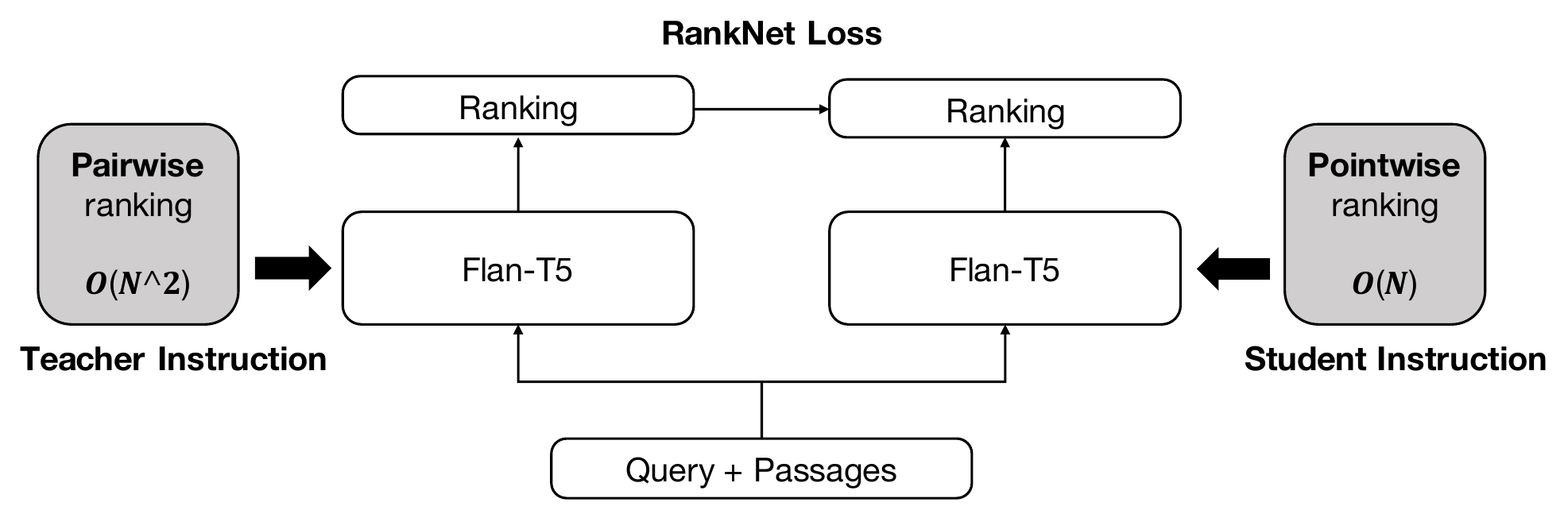} 
\caption{An overview of the proposed instruction distillation approach. Instruction distillation distills the abilities harvested from complex instruction techniques into a model that is more efficient with simple instruction techniques.}
\label{fig:distill}
\end{figure}

The key idea of Instruction Distillation is to distill the ability obtained from the complex but effective instruction technique (e.g., pairwise ranking instruction) into a model that is more efficient with the simple instruction technique (e.g., pointwise ranking instruction).
Figure~\ref{fig:distill} shows an overview of the propose instruction distillation approach.
We denote the sources of relevance scores or ranking results with superscripts ${}^\text{t}$ and ${}^\text{s}$ for teacher instruction and simplified student instruction, respectively. 
Our method unfolds in three stages: (1) Candidate generation, (2) Teacher inference, and (3) Student learning.

\begin{itemize}
    \item \textbf{Candidate generation.} 
    Suppose we have a dataset comprising a set of queries $Q$ and a corresponding set of items $\mathcal{D}$. It is worth mentioning that none of the queries require a labeled item.
    For a query $q \in Q$, an unsupervised retriever (e.g., BM25) is employed to fetch $n$ potentially relevant candidate samples $D = (d_1, d_2, \ldots, d_n)$ from the item set $\mathcal{D}$.

    \item \textbf{Teacher inference.} 
    Then, LLMs with costly pairwise ranking are employed as the \emph{teacher} models to re-rank the candidate set $D = (d_1, d_2, \ldots, d_n)$ corresponding to each query $q$. 
    To adopt the pairwise method, the $n$ items are juxtaposed in pairs, resulting in $n(n-1)$ ordered tuples $(d_i, d_j)$ where $i \neq j$. The model then scores the relevance of $d_i$ and $d_j$ to the given query $q$ using Eq.~(\ref{eq:pair}).  Based on these scores, each document $d_i$ is assigned a rank $r_i^{\text{t}}$ for every query $q$.

    \item \textbf{Student learning.} 
    In this phase, the pointwise ranking model serves as the \emph{student}.
    To leverage the ranking lists $r_i^\text{t}$ generated by the teacher, we employ the RankNet loss~\citep{Burges2005LearningTR} to optimize the student model. 
    RankNet is a pairwise loss function that measures the accuracy of relative ordering between items:
    \begin{equation*}
        \mathcal{L} = \sum_{i=1}^{n} \sum_{j=1}^{n} \mathbb{1}_{r_i^{\text{t}} < r_j^{\text{t}}} \log (1 + \exp (s_i^{\text{s}} - s_j^{\text{s}}))
    \end{equation*}
    Unlike other loss functions that utilize a sparse signal, the RankNet loss offers a richer transfer of ranking information from the teacher to the student.
\end{itemize}

After the instruction distillation process, the pointwise instruction technique is utilized during the inference stage. See Appendix~\ref{chap:appendix-A} for more details about the prompts. 

\section{Experimental Setup}
In order to comprehensively validate the effectiveness of the proposed method.
We conduct experiments on a variety of IR tasks, including both the text-based \textit{passage re-ranking} task and the item-based \textit{conversational recommendation} task. 

For passage re-ranking, the training data contain 10K queries sampled from the MS MARCO dataset ~\citep{Campos2016MSMA}. 
Each query is then paired with the top 10 documents retrieved by BM25. 
The trained models are evaluated on subtasks of TREC ~\citep{Craswell2020OverviewOT} benchmarks and BEIR ~\citep{Thakur2021BEIRAH} benchmarks.
NDCG@{1, 5, 10} are chosen as the metrics.

For conversational recommendation, we use the ReDial dataset ~\citep{li2018conversational}, which is a movie recommendation task based on conversation logs between the user and the recommender.
The trained models are then evaluated on the official test set.
For this setting, Acc@1 is adopted as the metric.

\subsection{Datasets}
\header{TREC}~\citep{Campos2016MSMA} is a widely used benchmark dataset in IR research. We use the test sets of the 2019 and 2020 competitions. TREC-DL19 and TREC-DL20 are both derived from MS MARCO datasets with human-generated labels. Each query is paired with 100 retrieved documents retrieved by BM25. They share the same format. TREC-DL19 contains 43 test queries, and TREC-DL20 contains 54 test queries.

\header{BEIR}~\citep{Thakur2021BEIRAH} consists of diverse retrieval tasks and domains.
We choose eight tasks in BEIR to evaluate the models:
(1) Covid retrieves scientific articles for COVID-19 related questions.
(2) NFCorpus is a bio-medical IR data.
(3) Touche is a argument retrieval datasets.
(4) {DBPedia} retrieves entities from DBpedia corpus.
(5) {SciFact} retrieves evidence for claims verification.
(6) {Signal} retrieves relevant tweets for a given news title.
(7) {News} retrieves relevant news articles for news headlines.
(8) {Robust04} evaluates poorly performing topics.
The evaluation results are averaged over the eight datasets.

\header{Redial} (Recommendation Dialogues)~\citep{Li2018TowardsDC} is an annotated conversational movie recommendation dataset, where users recommend movies to each other.

\subsection{Baselines}
To compare our methods with existing unsupervised and supervised methods, we choose widely applied methods as below:
\begin{itemize}
    \item \textbf{BM25} is an unsupervised, based on weighted term frequency. It is one of most the commonly adopted retrieval methods.
    \item \textbf{RankGPT}~\citep{Sun2023IsCG} is a listwise permutation generation approach based on \code{gpt-3.5-turbo} and \code{gpt-4}.
    \item \textbf{Relevance Gerneration}~\citep{Sachan2022ImprovingPR} is a pointwise ranking method based on FLAN-T5.
    \item \textbf{PRP}~\citep{Qin2023LargeLM} is a pairwise ranking ranking method based on FLAN-T5.
    \item \textbf{MonoT5}~\citep{Sachan2022QuestionsAA} is pointwise ranking method based on T5 models and is supervised trained on MS MARCO. 
    \item \textbf{Cohere Rerank} is a commercial text ranking system developed by Cohere\footnote{\url{https://cohere.com/rerank}}.
\end{itemize}

\begin{table*}[!t]
\centering
\small
\setlength\tabcolsep{4.5pt}
\caption{Results on TREC-DL19 and TREC-DL20 by re-ranking top-100 passages retrieved by BM25. Sec/Q indicates the average time in seconds to the re-rank 100 passages for a query. Best performing unsupervised and overall system(s) are marked bold.}
\label{table:trec}

\begin{tabular}{l l c c c}

\toprule

& & & \textbf{DL19}
& \textbf{DL20}
\\

Method  & LLM & Sec/Q & nDCG@1/5/10 & nDCG@1/5/10\\

\midrule

BM25 & -- & --
& 54.26 / 52.78 / 50.58 
& 57.72 / 50.67 / 47.96
\\
\midrule
\multicolumn{5}{c}{\textbf{Supervised LLMs Methods}}\\
\midrule

monoT5 & T5-Base & 0.12
& 79.84 / 73.77 / 71.48
& 77.47 / 69.40 / 66.99
\\

monoT5 & T5-XL & 1.30
& 79.07 / 73.74 / 71.83
& 80.25 / 72.32 / 68.89
\\

Cohere Rerank & \code{english-v2.0} & --
& 77.13 / 76.17 / 73.22
& 79.32 / 71.00 / 67.08
\\

\midrule
\multicolumn{5}{c}{\textbf{Unsupervised LLMs Methods}}\\
\midrule


RankGPT & \code{gpt-3.5-turbo} & --
& 82.17 / 71.15 / 65.80
& 79.32 / 66.76 / 62.91
\\

RankGPT & \code{gpt-4} & --
& 82.56 / 79.16 / 75.59 
& 78.40 / 74.11 / 70.56
\\

\midrule

Relevance Generation & FLAN-T5-Base & 0.12
& 58.13 / 48.52 / 47.43
& 55.25 / 50.35 / 48.32
\\

PRP (Allpair) & FLAN-T5-Base & 21.51
& 51.16 / 53.44 / 51.45
& 53.40 / 48.61 / 48.36
\\

\rowcolor{Gray} Instruction Distillation & FLAN-T5-Base & 0.12
& 59.69 / 60.21 / 57.30
& 63.27 / 55.50 / 53.09
\\

\midrule

Relevance Generation & FLAN-T5-Large & 1.10
& 43.41 / 47.65 / 48.41
& 40.43 / 45.19 / 46.67
\\

PRP (Allpair) & FLAN-T5-Large & 49.19
& 74.03 / 69.00 / 66.58
& 68.21 / 64.63 / 61.51
\\

\rowcolor{Gray} Instruction Distillation & FLAN-T5-Large & 1.10
& 74.33 / 74.18 / 69.81
& 72.84 / 65.59 / 62.80
\\

\midrule

Relevance Generation & FLAN-T5-XL & 1.30
& 50.00 / 54.33 / 52.85
& 45.37 / 48.56 / 49.07
\\

PRP (Allpair) & FLAN-T5-XL & 112.12
& 77.91 / 73.46 / 70.58
& 76.85 / 69.58 / 67.21 
\\

\rowcolor{Gray} Instruction Distillation & FLAN-T5-XL & 1.30
& 79.85 / 75.15 / 71.92
& 81.17 / 72.08 / 69.29
\\





\bottomrule
\end{tabular}

\end{table*}

\subsection{Implementation Details}

\header{Passage Re-Ranking Task.}
Following \citet{Sun2023IsCG}, we sample 10K queries from the MS MARCO training set. 
Utilizing BM25 as the candidate generator, we retrieve 10 passages for each query.
Our BM25 implementation is derived from BM25Okapi as presented in RankBM25~\citep{trotman2014improvements}. 
Prior to retrieval, we ensure that stopwords are eliminated. 
In implementing the pairwise prompting strategy, each query's 10 passages are juxtaposed in pairs, leading to the generation of 90 ordered passage pairs. 
The teacher models are instructed to determine which document is more relevant to the query and subsequently produce the ranking results. The results are then used as the pseudo labels for pointwise instruction distillation. 
To harness the full potential of the ranking outcomes, we employ RankNet~\citep{Burges2005LearningTR}.

\header{Conversational Recommendation Task.}
For this task, we use the dialogue history as the query, the descriptions of movies as documents, and employ BM25 to fetch the top-5 movies into the candidate pool. 
Furthermore, following \citet{Hou2023LargeLM}, an additional 4 popular movies are incorporated into the candidate pool\footnote{The criterion for determining a movie's popularity is based on its frequency of mentions throughout the training dataset. Movies cited more than 200 times are classified as popular. The likelihood of selecting a popular movie is proportional to its representation in the overall popularity.}. This is done to simulate the inherent feature of popularity bias in recommendations~\citep{chen2023bias}.

\header{Training Details.}
Throughout the training phase, we employ the AdamW optimizer with a consistent learning rate of ${3e-5}$. We constrain the maximum input length to ${512}$ tokens. The training environment is 4 * A800-80G, with a batch size fixed at $32$. We train the model up to ${3}$ epochs. 
Our experiments are based on the FLAN-T5 family~\citep{chung2022scaling}, a suite of models which has been fine-tuned for various NLP tasks. Our experiments specifically leverage models such as FLAN-T5-XL (3B), FLAN-T5-Large (770M), and FLAN-T5-Base (220M).

The prompts used can be seen in Appendix~\ref{chap:appendix-A}.

\section{Experimental Results}
\subsection{Results on Passage Re-Ranking Tasks}

The experimental results on TREC and BEIR datasets are presented in Table~\ref{table:trec} and Table~\ref{table:beir} respectively. Based on these results, we draw the following observations:

Firstly, when compared with previous unsupervised LLM prompting strategies, our instruction-distilled models' inference speed aligns with that of the \textit{Relevance Generation} method, and it is notably over $100\times$ faster than the PRP method. Moreover, the performance of our approach using FLAN-T5-XL and FLAN-T5-Large surpasses both the Relevance Generation and PRP methods with the same LLMs.

Secondly, the instruction-distilled models yield results akin to their supervised counterparts but with reduced annotation requirements. Specifically, our instruction-distilled FLAN-T5-XL model achieves nDCG@10 of 71.92 and 69.29 on TREC-DL19 and TREC-DL20, respectively, either matches or surpasses the performance of the supervised monoT5 of equivalent parameter size.

Lastly, the instruction-distilled models always perform superior to their teachers. For example, the distilled models of all different model sizes perform better than their PRP teachers. This can be attributed to the fact that unspecialized teacher models might produce unstable outputs. After distillation on task-related data, student models are able to strictly follow the given instructions, generating more reliable outputs. This specialization phase significantly enhances both the efficiency and performance of all involved models.

Similar findings can be observed on the BEIR dataset.

\begin{table}[!t]
\centering
\small
\setlength\tabcolsep{1.8pt}
\caption{Results (nDCG@10) on BEIR.}
\label{table:beir}
\begin{tabular}{ll cccccccc c}

\toprule

Method & LLM & Covid & NFC. & Touche & DBP. & SciFact & Signal & News & Robust04 & \textbf{Avg.} 
\\
\midrule

BM25 & -- & 59.47 & 30.75 & 44.22 & 31.80 & 67.89 & 33.05 & 39.52 & 40.70 & 43.42 \\
monoT5 & T5-Base & 78.34 & 37.38 & 30.82 & 42.42 & 73.40 & 31.67 & 46.83 & 51.72 & 49.07 \\
monoT5 & T5-XL & 80.71 & 38.97 & 32.41 & 44.45 &  76.57 & 32.55 & 48.49 & 56.71 & 51.36 \\
Cohere Rerank & \code{english-v2.0} & 81.81 & 36.36 & 32.51 & 42.51 & 74.44 & 29.60 & 47.59 & 50.78 & 49.45 \\
RankGPT & \code{gpt-3.5-turbo} & 76.67 & 35.62 & 36.18 & 44.47 & 70.43 & 32.12 & 48.85 & 50.62 & 49.37 \\
RankGPT & \code{gpt-4} & 85.51 & 38.47 & 38.57 & 47.12 & 74.95 & 34.40 & 52.89 & 57.55 & 53.68\\

\midrule
Ours & FLAN-T5-XL & 80.96 	& 38.25 	& 30.97 	& 45.09 	& 75.66 	& 32.45 	& 49.21 	& 56.64 	& 51.15 
\\
Ours & FLAN-T5-Large & 79.95 	& 35.41 	& 30.25 	& 45.22 	& 71.22 	& 30.80 	& 44.52 	& 49.22 	& 48.32 
\\
Ours & FLAN-T5-Base & 69.11 	& 30.51 	& 24.10 	& 32.15 	& 36.92 	& 28.84 	& 31.98 	& 37.65 	& 36.41 
\\

\bottomrule
\end{tabular}
\end{table}

\subsection{Results on Conversational Recommendation Tasks}
Understanding user preferences from dialogue history presents a greater challenge than merely ranking relevance based on a specified query. Despite this, our method demonstrates noteworthy results, which are summarized in Table~\ref{table:redial}.

Firstly, our method achieves the best results among all the unsupervised methods. Specifically, our distillation technique outperforms other methods across all scales in terms of Acc@1 metrics. The FLAN-T5-XL distilled model achieves a peak value of $24.93\%$ on Acc@1, outperforming all other unsupervised models.

Secondly, when compared with the teacher model, the student model exhibits either comparable or superior performance. 
The teacher model, employing FLAN-T5-XL with PRP techniques, posts an Acc@1 of $20\%$. In contrast, the distilled model with equivalent parameter size achieves an impressive $24.93\%$ in terms of Acc@1. Meanwhile, the Large model, with less than a third of the teacher model's parameters, records a close Acc@1 score of $19.71\%$.

\begin{table}[h]
\centering
\small
\setlength\tabcolsep{7pt}
\caption{Results (Acc) on REDIAL.}
\label{table:redial}
\begin{tabular}{llcc }

\toprule

Method & LLM & Sec/Q &  Acc
\\
\midrule

Random & -- & -- & 10.77\\
Popularity & -- & -- &  7.69\\
BM25 & -- & -- &  8.62\\
\midrule
\multicolumn{3}{l}{\textbf{Unsupervised LLMs Methods}}\\
\midrule
Listwise Ranking & T5-XL & 0.02 & 16.92\\
Pairwise Ranking & T5-XL &  7.90 & 20.00\\
Pointwise Ranking & T5-XL &  1.44 & 12.00\\
\rowcolor{Gray} Instruction Distillation & T5-XL & 1.44 & 24.93\\
\midrule
Listwise Ranking & T5-Large & 0.01 & 13.85\\
Pairwise Ranking & T5-Large & 3.06 & 16.62\\
Pointwise Ranking & T5-Large & 0.49 & 8.00\\
\rowcolor{Gray} Instruction Distillation & T5-Large & 0.49 & 19.71\\
\midrule
Listwise Ranking & T5-Base & 0.01 & 1.54\\
Pairwise Ranking & T5-Base & 1.00 & 13.69\\
Pointwise Ranking & T5-Base & 0.18 & 10.77\\
\rowcolor{Gray} Instruction Distillation & T5-Base & 0.18 & 15.07\\
\bottomrule
\end{tabular}
\end{table}

Lastly, there is a notable improvement in the performance metrics of all the distilled models after instruction distillation. For instance, the FLAN-T5-XL model, when used with the pointwise prompt, only marginally surpasses the random recommendation. However, after the proposed instruction distillation process, its Acc@1 nearly doubles. A similar improvement is observed for FLAN-T5-Large, with its Acc@1 soaring from $8\%$ to $19.71\%$. Even though the increase might not seem substantial due to the model's capacity, it represents a growth of over $5\%$.

\subsection{Analytical Experiments}

\begin{figure}[h]
\centering
\includegraphics[width=0.6\columnwidth]{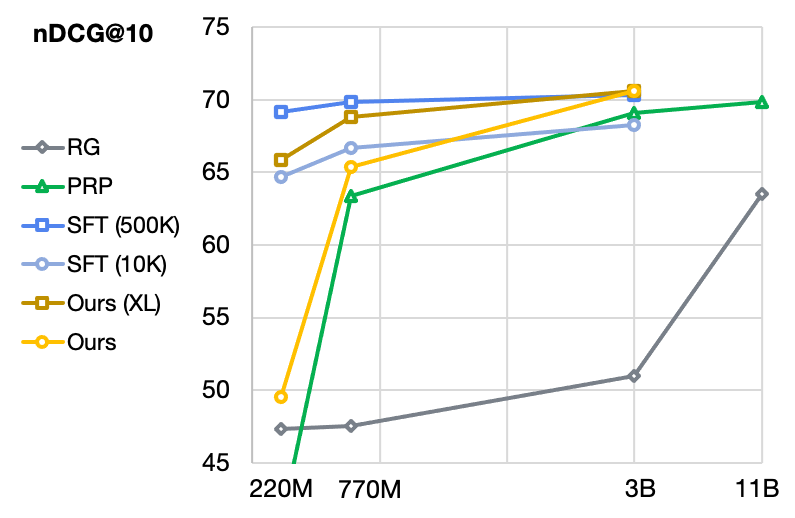} 
\caption{Compare the proposed method with baselines in terms of model size. We can see that our methods (denoted by yellow line) outperform supervised finetuning (SFT) methods when the number of parameters exceeds 3B.}
\label{fig:vs-sft}
\end{figure}

To gain deeper insights into the impact of model size and training signal, we carried out an analytical experiment. The results are depicted in Figure~\ref{fig:vs-sft}. Several key observations can be made from these results:
(1) Instruction distillation models, represented by the yellow line in the figure, outperform the state-of-the-art supervised system, monoT5 (or SFT (500K), illustrated by the blue line), when the model size surpasses 3B. Moreover, our approach consistently exceeds the performance of earlier zero-shot LLM methods, namely RG and PRP, across all scales.
(2) Distilling from larger models can enhance the performance of their smaller counterparts. 
As evidenced by our results labeled ``Ours (XL)'' in Figure~\ref{fig:vs-sft} -- which captures the process of distilling the predictions from FLAN-T5-XL to smaller models -- it becomes clear that instruction distillation from larger models invariably boosts the capabilities of smaller ones.
(3) Given the same training data size, our approach, which distilling from FLAN-T5-XL (referred to as ``Ours (XL)'' in Figure~\ref{fig:vs-sft}) and is unsupervised, significantly outperforms its supervised counterpart (referred to as ``SFT (10k)'' in Figure~\ref{fig:vs-sft}). This finding shows the promising potential of leveraging LLMs as data labelers in ranking tasks.

\section{Conclusion}
This paper proposes instruction distillation, an unsupervised method that distills LLMs' abilities uncovered by complex instructions into the same model but with simpler instructions. 
This method significantly improves the efficiency and stability of LLMs, which is very friendly for industrial application deployment. 
Our experimental results on passage ranking and conversational recommendation verify the effectiveness of the proposed method.
With our method, the efficiency of the models is significantly improved. A 10--100$\times$ increase in efficiency can be observed when compared to comparable unsupervised methods.

\bibliographystyle{acl_natbib}
\bibliography{sample-base}


\onecolumn
\appendix

\section{Prompts} \label{chap:appendix-A}

\subsection{Passage Ranking}
\newtcolorbox{promptbox}[1]{colback=Gray!5!white,colframe=Gray!95!black,fonttitle=\bfseries,title=#1}

\begin{promptbox}{Pointwise Ranking Prompt}
Question: Given a query ``\code{\{\{query\}\}}'', Is the following passage relevant to the query?\\

Passage : \code{\{\{passage\}\}}\\

If it is relevant answer Yes, else answer No. \\

Answer:
\end{promptbox}

\begin{promptbox}{Pairwise Ranking Prompt}
Question: Given a query ``\code{\{\{query\}\}}'', which of the following two passages is more relevant to the query?\\

passage A: \code{\{\{passage\_A\}\}}\\

passage B: \code{\{\{passage\_B\}\}}\\

Output the identifier of the more relevant passage. The answer must be passage A or passage B.\\

Answer:
\end{promptbox}


\subsection{Conversational Recommendation}

\newtcolorbox{enbox}[1]{colback=Gray!5!white,colframe=Gray!95!black,fonttitle=\bfseries,title=#1}

\begin{enbox}{Pointwise Ranking Prompt}
Question: Given the conversation history between the recommender and the user:\\

\code{\{\{query\}\}}\\

Based on the user's preference, is the following movie suitable to the user?\\

Movie: {\code{\{\{movie\}\}}}\\

The answer must be Y or N. Give the answer after Answer: .
\end{enbox}

\begin{enbox}{Pairwise Ranking Prompt}
Question: Given the conversation history between the recommender and the user:\\

\code{\{\{query\}\}}\\

Based on the user's preference, which of the following two movies is more suitable to the user?\\

Movie A: {\code{\{\{movie\_A\}\}}}\\

Movie B: {\code{\{\{movie\_B\}\}}}\\

The answer must be A or B. Give the answer after the Answer: .
\end{enbox}

\begin{enbox}{Listwise Ranking Prompt}
Question: Given the conversation history between the recommender and the user: \\

\code{\{\{query\}\}}\\

Based on the user's preference, which of the following movies is the most suitable for the user?\\

[1]: {\code{\{\{movie\_1\}\}}}\\

[2]: {\code{\{\{movie\_2\}\}}}\\

...\\

Answer the question with the number of the movie. The answer will include one and only one number. Give the answer after Answer: . 
\end{enbox}

\end{document}